\documentclass[envcountsect,runningheads,10pt]{llncs}

\usepackage[letterpaper]{geometry}

\usepackage{url,verbatim}
\usepackage{algorithm}
\usepackage{algorithmic}

\usepackage{amsmath,amssymb}

\newcommand{\ignore}[1]{}

\newcommand{\F}{\mathbb{F}}
\newcommand{\f}{\mathcal{F}}
\newcommand{\Z}{\mathbb{Z}}
\newcommand{\p}{\mathfrak{p}}
\renewcommand{\v}{\mathbf{v}}
\newcommand{\z}{\mathbf{z}}
\renewcommand{\a}{\mathfrak{a}}
\renewcommand{\O}{\mathcal{O}}
\renewcommand{\b}{\mathfrak{b}}
\renewcommand{\aa}{\mathbf{a}}

\newcommand{\Smooth}{\mathcal{S}}
\newcommand{\Primes}{\mathcal{P}}
\newcommand{\nil}{\texttt{nil}}
\newcommand{\union}{\cup}

\DeclareMathOperator{\End}{End}

\DeclareMathOperator{\Cl}{Cl}
\DeclareMathOperator{\Ell}{Ell}

\newtheorem{prop}[theorem]{Proposition}
\newtheorem{cor}[theorem]{Corollary}
\spnewtheorem{rem}[theorem]{Remark}{\itshape}{\rmfamily}
\spnewtheorem{defn}[theorem]{Definition}{\bfseries}{\rmfamily}

\usepackage{colonequals}
\newcommand{\step}[1]{Step~\ref{step:#1}}
\newcommand{\steprange}[2]{Steps~\ref{step:#1}--\ref{step:#2}}
\newcommand{\alg}[1]{Algorithm~\ref{alg:#1}}
\newcommand{\lem}[1]{Lemma~\ref{lem:#1}}

\newcommand{\thm}[1]{Theorem~\ref{thm:#1}}
\newcommand{\corol}[1]{Corollary~\ref{cor:#1}}
\renewcommand{\sec}[1]{Section~\ref{sec:#1}}
\newcommand{\rmrk}[1]{Remark~\ref{rem:#1}}
\newcommand{\poly}{\mathop{\mathrm{poly}}}
\newcommand{\spn}{\mathop{\mathrm{span}}}
\newcommand{\defeq}{\colonequals}
\renewcommand{\>}{\rangle}
\newcommand{\<}{\langle}
\newcommand{\floor}[1]{\lfloor{#1}\rfloor}

\newcommand{\Floor}[1]{\left\lfloor{#1}\right\rfloor}
\newcommand{\Ceil}[1]{\left\lceil{#1}\right\rceil}

\newcommand{\mixrad}[1]{\mu(#1)}
\usepackage{graphicx}
\newcommand{\varstar}{{\rotatebox[origin=c]{180}{$\scriptstyle\star$}}}
\newcommand{\inrand}{\in_{\mathrm{R}}}

\begin{document}

\title{Constructing elliptic curve isogenies \\ in quantum subexponential time}
\author{Andrew M.\ Childs\inst{1,2} \and David Jao\inst{1} \and Vladimir Soukharev\inst{1}}
\institute{
Department of Combinatorics and Optimization,\\
University of Waterloo, Waterloo, Ontario, N2L 3G1, Canada \and
Institute for Quantum Computing,\\
University of Waterloo, Waterloo, Ontario, N2L 3G1, Canada\\[4pt]
\email{\{amchilds,djao,vsoukhar\}@math.uwaterloo.ca}
}

\maketitle

\begin{abstract}
  Given two elliptic curves over a finite field having the same
  cardinality and endomorphism ring, it is known that the curves admit
  an isogeny between them, but finding such an isogeny is believed to
  be computationally difficult. The fastest known classical algorithm
  takes exponential time, and prior to our work no faster quantum
  algorithm was known. Recently, public-key cryptosystems based on the
  presumed hardness of this problem have been proposed as candidates
  for post-quantum cryptography. In this paper, we give a
  subexponential-time quantum algorithm for constructing isogenies,
  assuming the Generalized Riemann Hypothesis (but with no other
  assumptions). Our algorithm is based on a reduction to a hidden
  shift problem, together with a new subexponential-time algorithm for
  evaluating isogenies from kernel ideals (under only GRH), and
  represents the first nontrivial application of Kuperberg's quantum
  algorithm for the hidden shift problem.  This result
  suggests that isogeny-based cryptosystems may be uncompetitive with
  more mainstream quantum-resistant cryptosystems such as
  lattice-based cryptosystems.
  
\end{abstract}

\section{Introduction}\label{sec:intro}

We consider the problem of constructing an isogeny between two given
isogenous ordinary elliptic curves defined over a finite field $\F_q$
and having the same endomorphism ring. (Two such curves are called
\emph{horizontally isogenous}.) This problem has led to several
applications in elliptic curve cryptography, both constructive and
destructive. The fastest known probabilistic algorithm for solving
this problem is the algorithm of Galbraith and Stolbunov~\cite{gs},
based on the work of Galbraith, Hess, and Smart~\cite{GHS}. Their
algorithm is exponential, with a worst-case (and average-case) running
time roughly proportional to $\sqrt[4]{q}$.

Although quantum attacks are known against several cryptographic
protocols of an algebraic nature \cite{DHI03,hallgren-pell,shor},
until now there has been no nontrivial quantum algorithm for
constructing isogenies. The difficulty of this problem has led to
various constructions of public-key cryptosystems based on finding
isogenies, beginning with a proposal of Couveignes~\cite{hhs}.  More
recently, Rostovtsev and Stolbunov~\cite{rost} and
Stolbunov~\cite{stolbunov} proposed refined versions of these
cryptosystems with the specific aim of obtaining cryptographic protocols
that resist attacks by quantum computers.

In this work, we give a subexponential-time quantum algorithm for
constructing an isogeny between two given horizontally isogenous
elliptic curves, and show that the running time of our algorithm is
bounded above by $L_q(\frac12,\frac{\sqrt3}{2})$ under (only) the
Generalized Riemann Hypothesis (GRH). This result raises serious
questions about the viability of isogeny-based cryptosystems in the
context of quantum computers. At present, isogeny-based
cryptosystems are not especially attractive since their performance is
poor compared to other quantum-resistant cryptosystems, such as
lattice-based cryptography~\cite{ntru}. Nevertheless, they represent a
distinct family of cryptosystems worthy of analysis (for reasons of
diversity if nothing else, given the small number of quantum-resistant
public-key cryptosystem families available~\cite{qrsurvey}).  Since
isogeny-based cryptosystems already perform poorly at moderate
security levels~\cite[Table 1]{stolbunov}, any improved attacks such
as ours would seem to disqualify such systems from consideration in a
post-quantum world.

\subsection{Contributions}

Our first main contribution, described in \sec{quantum}, is a
reduction from the problem of isogeny construction to the abelian
hidden shift problem.  While a connection between isogenies and hidden
shifts was noted previously by Stolbunov \cite{stolbunov}, we observe
that the reduction gives an \emph{injective} hidden shift problem.
This allows us to apply an algorithm of Kuperberg \cite{kuperberg} to
solve the hidden shift problem using a subexponential number of
queries to certain functions.  This reduction constitutes the first
nontrivial application of Kuperberg's algorithm outside of the
black-box setting.

The reduction to the hidden shift problem alone does not immediately
give a subexponential-time algorithm for computing isogenies, because
one must consider the time required to compute the hiding functions.
Indeed, prior to our work there was no known subexponential-time
algorithm to evaluate these functions.  Our second main contribution,
described in \sec{main}, is a subexponential-time (classical)
algorithm to compute the isogeny star operator, which is defined as a
certain action of an ideal class group on a set of elliptic
curves. In this way we can compute the hiding functions in
subexponential time and thus obtain a subexponential-time reduction to
the hidden shift problem.  Unlike previous algorithms for isogeny
computation~\cite{galbraith,GHS,jv}, our runtime analysis assumes
only GRH, whereas all previous subexponential-time algorithms for
isogeny problems have required additional heuristic assumptions.  We
achieve this improvement using expansion properties of a certain Cayley
graph \cite{jnt-2009}. The same idea can also be used to obtain
subexponential algorithms (under only GRH) for evaluating isogenies (see
\rmrk{evaluate}). In addition, Bisson~\cite{bisson-2011} has shown
that our method yields a subexponential algorithm for computing
endomorphism rings of ordinary elliptic curves under GRH.

Kuperberg's algorithm for the abelian hidden shift problem uses
superpolynomial space (i.e., a quantum computer with superpolynomially
many qubits), so the same is true of the most straightforward version
of our algorithm.  Since it is difficult to build quantum computers
with many qubits, this feature could limit the
applicability of our result.  However, we also obtain an algorithm
using polynomial space by taking advantage of an alternative approach
to the abelian hidden shift problem due to Regev
\cite{regev-polyspace}.  Regev only explicitly considered the case of
the hidden shift problem in a cyclic group whose order is a power of
$2$, and even in that case did not compute the constant in the
exponent of the running time.  We fill both of these gaps in our work,
showing that the hidden shift problem in any finite abelian group $A$
can be solved in time $L_{|A|}(\frac12,\sqrt2)$ by a quantum computer
using only polynomial space.  Consequently, we give a polynomial-space
quantum algorithm for isogeny construction using time
$L_q(\frac12,\frac{\sqrt3}{2}+\sqrt2)$.  The group relevant to isogeny
construction is not always cyclic, so the extension to general abelian 
groups is necessary for our application.

\subsection{Related work}\label{sec:related}

Our algorithm for evaluating the isogeny star operator is based on
reducing an ideal modulo principal ideals to obtain a smooth ideal.
This idea is originally due to Galbraith, Hess, and Smart~\cite{GHS}.
Br\"oker, Charles, and Lauter~\cite{BCL} and Jao and
Soukharev~\cite{jv} also use this idea to give algorithms for
evaluating isogenies. Bisson and Sutherland~\cite{bisson-2009} use a
similar smoothing technique to compute endomorphism rings in
subexponential time. We stress that, with the exception of~\cite{BCL},
which is restricted in scope to small discriminants, all the results
mentioned above make heuristic assumptions of varying
severity~\cite[\S 4]{bisson-2009}\cite[p. 37]{GHS}\cite[p.  224]{jv}
in addition to the Generalized Riemann Hypothesis in the course of
proving their respective runtime claims. Our work is the first to
achieve provably subexponential running time with no heuristic
assumptions other than GRH. In practice, the heuristic algorithms
in~\cite{bisson-2009} and~\cite{jv} run slightly faster than our
algorithms in Section~\ref{sec:main}, because they make use of an
optimized exponent distribution (originating from~\cite{bisson-2009})
that minimizes the number of large degree isogenies appearing in
the smooth factorization. Our work does not use this optimization,
because doing so would reintroduce the need for additional heuristic
assumptions.

An alternative approach to computing isogenies, given in
Couveignes~\cite[p. 11]{hhs} and Stolbunov~\cite[p. 227]{stolbunov},
is to treat the class group as a $\Z$-module and use lattice basis
reduction to compute the isogeny star operator. In practice, the
lattice-based approach works well for moderate parameter
sizes. However, since it amounts to solving the closest vector
problem, the method asymptotically requires exponential time
(even with known quantum algorithms), and thus is slower than our
approach.

\section{Isogenies}\label{sec:background}

For general background on elliptic curves, we refer the reader to
Silverman~\cite{silverman}.

Let $E$ and $E'$ be elliptic curves defined over a field $F$. An
\emph{isogeny} $\phi\colon E \to E'$ is an algebraic morphism
satisfying $\phi(\infty) = \infty$. The \emph{degree} of an isogeny is
its degree as an algebraic map. The \emph{endomorphism ring} $\End(E)$
is the set of isogenies from $E(\bar{F})$ to itself. This set forms a
ring under pointwise addition and composition.

When $F$ is a finite field, the rank of $\End(E)$ as a $\Z$-module is
either $2$ or $4$. We say $E$ is \emph{supersingular} if the rank is
$4$, and \emph{ordinary} otherwise. A supersingular curve cannot be
isogenous to an ordinary curve.
Most elliptic curves are ordinary (in particular, supersingular curves
have density zero~\cite{se3}), and most current proposals for
isogeny-based cryptography (including all published isogeny-based
public-key cryptosystems) use ordinary curves.  Thus, in this paper
we restrict our attention to ordinary elliptic curves.  It remains an
interesting open problem to study cryptographic applications of
isogenies between supersingular curves and to better understand the
computational difficulty of computing such isogenies, but we do not
address this issue.

Over a finite field $\F_q$, two elliptic curves $E$ and $E'$ are
isogenous if and only if $\#E(\F_q) = \#E'(\F_q)$~\cite{tate}.  The
endomorphism ring of an ordinary elliptic curve over a finite field is
an imaginary quadratic order $\O_\Delta$ of discriminant $\Delta < 0$.
The set of all isomorphism classes (over $\bar{\F}_q$) of isogenous
curves with endomorphism ring $\O_\Delta$ is denoted
$\Ell_{q,n}(\O_\Delta)$, where $n$ is the cardinality of any such
curve. We represent elements of $\Ell_{q,n}(\O_\Delta)$ by taking the
$j$-invariant of any representative curve in the isomorphism class.

An isogeny between two curves having the same endomorphism ring is called
a \emph{horizontal} isogeny~\cite{volcano}. Likewise, we say that two
isogenous curves are \emph{horizontally isogenous} if their
endomorphism rings are equal.  Any separable horizontal isogeny
$\phi\colon E \to E'$ between curves in $\Ell_{q,n}(\O_\Delta)$ can be
specified, up to isomorphism, by giving $E$ and
$\ker\phi$~\cite[III.4.12]{silverman}.  The kernel of an isogeny, in
turn, can be represented as an ideal in
$\O_\Delta$~\cite[Thm. 4.5]{waterhouse}. Denote by $\phi_\b\colon E
\to E_\b$ the isogeny corresponding to an ideal $\b$ (keeping in mind
that $\phi_\b$ is only defined up to isomorphism of $E_\b$). Principal
ideals correspond to isomorphisms, so any other ideal equivalent to
$\b$ in the ideal class group $\Cl(\O_\Delta)$ of $\O_\Delta$ induces
the same isogeny, up to isomorphism~\cite[Thm.
  3.11]{waterhouse}. Hence one obtains a well-defined group action
\begin{align*}
*\colon \Cl(\O_\Delta) \times \Ell_{q,n}(\O_\Delta) &\to
\Ell_{q,n}(\O_\Delta)\\
[\b] * j(E) &= j(E_\b)
\end{align*}
where $[\b]$ denotes the ideal class of $\b$. This group action, which
we call the \emph{isogeny star operator}, is free and
transitive~\cite[Thm. 4.5]{waterhouse}, and thus
$\Ell_{q,n}(\O_\Delta)$ forms a principal homogeneous space over
$\Cl(\O_\Delta)$.

\subsection*{Isogeny graphs under GRH}

Our runtime analysis in Section~\ref{sec:main} relies on the following
result of \cite{jnt-2009} which states, roughly, that random short products of small
primes in $\Cl(\O_\Delta)$ yield nearly uniformly random elements of
$\Cl(\O_\Delta)$, under GRH.

\begin{theorem}\label{thm:grh-graphs}
  Let $\O_\Delta$ be an imaginary quadratic order of discriminant
  $\Delta < 0$ and conductor $c$.
    Set $G =
  \Cl(\O_\Delta)$. Let $B$ and $x$ be real numbers satisfying $B > 2$
  and $x \geq (\ln |\Delta|)^B$. Let $S_x$ be the multiset $A \union
  A^{-1}$ where
\[
A = \{[\p] \in G : \gcd(c,\p) = 1 \text{ and } N(\p)
\leq x \text{ is prime} \}
\]
with $N(\p)$ denoting the norm of $\p$.
Then, assuming GRH, there exists a positive absolute constant $C > 1$,
depending only on $B$, such that for all $\Delta$, a random walk of
length
\[
t \geq C \frac{\ln|G|}{\ln \ln |\Delta|}
\]
in the Cayley graph $\operatorname{Cay}(G, S_x)$ from
any starting vertex lands in any fixed subset $S \subset
G$ with probability at least $\frac{1}{2} \frac{|S|}{|G|}$.
\end{theorem}

\begin{proof}
Apply Corollary~1.3 of~\cite{jnt-2009} with the parameters
\begin{itemize}
\item $K = $ the field of fractions of $\O_\Delta$
\item $G = \Cl(\O_\Delta)$
\item $q=|\Delta|$.
\end{itemize}
Following~\cite{cox}, we refer to $G = \Cl(\O_\Delta)$ as the
\emph{ring class group} of $\Delta$. Observe that by Remark~1.2(a)
of~\cite{jnt-2009}, Corollary~1.3 of~\cite{jnt-2009} applies to the
ring class group $G$, since ring class groups are
quotients of narrow ray class groups~\cite[p. 160]{cox}.  By
Corollary~1.3 of~\cite{jnt-2009}, Theorem~\ref{thm:grh-graphs} holds
for all sufficiently large values of $|\Delta|$, i.e., for all but
finitely many $|\Delta|$. To prove the theorem for all $|\Delta|$,
simply take a larger (but still finite) value of $C$.
\end{proof}

\begin{cor}\label{cor:grh-graphs}
  Theorem~\ref{thm:grh-graphs}
  still holds with the set $A$ redefined as
\[
A = \{[\p] \in G : \gcd(m\Delta,\p) = 1 \text{ and } N(\p)
\leq x \text{ is prime} \}
\]
where $m$ is any integer having at most $O(x^{1/2-\varepsilon}
\log|\Delta|)$ prime divisors.
\end{cor}
\begin{proof}
  The alternative definition of the set $A$ differs from the original
  definition by no more than $O(x^{1/2-\varepsilon} \log|\Delta|)$
  primes.  As stated in~\cite[p.  1497]{jnt-2009}, the contribution of
  these primes can be absorbed into the error term $O (x^{1/2} \log(x)
  \log(xq))$, and hence does not affect the conclusion of the theorem.
\end{proof}

\section{The group action inverse problem}\label{sec:crypto}

For a fixed discriminant $\Delta$, the \emph{vectorization}~\cite[\S
  2]{hhs} or \emph{group action inverse}~\cite[\S
  2.4]{stolbunov} problem is the problem of finding an ideal class
$[\b] \in \Cl(\O_\Delta)$ such that $[\b] * j(E) = j(E')$, given
$j(E)$ and $j(E')$. We refer to $[\b]$ as the
\emph{quotient} of $j(E)$ and $j(E')$. The computational infeasibility
of finding quotients in $\Ell_{q,n}(\O_\Delta)$ is a necessary
condition for the security of isogeny-based cryptosystems~\cite[\S
  3]{hhs}\cite[\S 7]{stolbunov}.  In the remainder of this
paper, we present our subexponential algorithm for evaluating
quotients in $\Ell_{q,n}(\O_\Delta)$ on a quantum computer.

A notable property of isogeny-based cryptosystems is that they do not
require the ability to evaluate the isogeny star operator efficiently
on arbitrary inputs.  It is enough to sample from random smooth ideals
(for which $*$ can be evaluated efficiently) when performing
operations such as key generation~\cite[\S 5.4]{hhs}\cite[\S
  6.2]{stolbunov}. However, to attack these cryptosystems using our
approach, we \emph{do} require the ability to evaluate the isogeny
star operator on arbitrary inputs. We turn to this problem in the next
section.

\section{Computing the isogeny star operator}\label{sec:main}

In this section, we describe a new classical (i.e., non-quantum)
algorithm to evaluate the isogeny star operator. All notation
is as in Section~\ref{sec:background}. Given an ideal class $[\b]$ in
$\Cl(\O_\Delta)$, and a $j$-invariant $E$ of an ordinary elliptic
curve of endomorphism ring $\O_\Delta$ over $\F_q$, we wish to
evaluate $[\b] * j(E)$. We define
\[\textstyle
  L_N(\frac12,c) \defeq \exp[(c + o(1)) \sqrt{\ln N \ln\ln N}].
\]
For convenience, we denote $L_{\max\{|\Delta|,q\}}(\frac12,c)$ by
$L(c)$.

In Section~\ref{subsec:runtime} we show that, under GRH, our algorithm
has a running time of $L_q(\frac12,\frac{\sqrt{3}}{2})$, which is
subexponential in the input size.
For clarity, we present our algorithms and analysis in
full instead of as ``patches'' to existing work.  We emphasize that
the basic structure of these algorithms appeared in prior work; our
main contribution is to the analysis, which is facilitated by small
changes to the algorithms. Specifically, Algorithm~\ref{alg:relation}
is based on~\cite[Algorithm 3]{jv}, which is in turn based on Seysen's
algorithm~\cite{seysen}; Algorithm~\ref{alg:main} is based
on~\cite[Algorithm 4.1]{BCL}.  Our bounds on $t$ in
Algorithm~\ref{alg:relation} are new, and allow us to prove the
crucial runtime bound
(Proposition~\ref{prop:relation-success-probability}).

\paragraph{Computing a relation.} Given an ideal class $[\b] \in
\Cl(\O_\Delta)$, Algorithm~\ref{alg:relation} produces a relation
vector $\z = (z_1,\ldots,z_f) \in \Z^f$ for $[\b]$, with respect to a
factor base $\f = \{\p_1,\ldots,\p_f\}$, satisfying $[\b] = \f^\z
\defeq \p_1^{z_1} \cdots \p_f^{z_f}$, with the additional property
(cf.\ Proposition~\ref{prop:relation-norm}) that the $L^1$-norm
$|\z|_1$ of $\z$ is less than $O(\ln|\Delta|)$ for some absolute
implied constant (here the $L^1$ norm of a vector denotes the sum of
the absolute values of its coordinates).
Algorithm~\ref{alg:relation}
is similar to Algorithm 11.2 in~\cite{buchmann}, except that we impose
a constraint on $|\v|_1$ in \step{rel-select} in order to keep $|\z|_1$
small, and (for performance reasons) we use Bernstein's algorithm
instead of trial division to find smooth elements.
We remark that Corollary~9.3.12 of~\cite{buchmann} together with the
restriction $C > 1$ in Theorem~\ref{thm:grh-graphs} implies that there
exists a value of $t$ satisfying the inequality in
Algorithm~\ref{alg:relation}.

\begin{algorithm}[t]
\caption{Computing a relation}
\label{alg:relation}
\begin{algorithmic}[1]
  \REQUIRE $\Delta$, $q$, $n$, $z$, $[\b]$, and an integer $t$
  satisfying $C\frac{\ln|{\Cl(\O_\Delta)}|}{\ln\ln|\Delta|}\leq t \leq
  C\ln|\Delta|$ where $C$ is the constant of
  Theorem~\ref{thm:grh-graphs}/Corollary~\ref{cor:grh-graphs}

\ENSURE A relation vector $\z \in \Z^f$ such that $[\b] = [\f^\z]$, or $\nil$

\STATE\label{step:factorbase} Compute a factor base consisting of
split primes; discard any primes dividing $qn$ to obtain a new factor
base $\f = \{\p_1, \p_2, \ldots, \p_f\}$

\STATE\label{step:rel-init1} Set $\Smooth \leftarrow \emptyset$,
$\Primes \leftarrow \{N(\p) : \p \in \f\}$

\STATE\label{step:rel-init2} Set $\ell \leftarrow L(\frac{1}{4z})$

\FOR{$i=0$ to $\ell$}\label{step:rel-startfor}

\STATE Select $\v \in \Z_{0..|\Delta|-1}^f$ uniformly at
random subject to the condition that $|\v|_1 = t$\label{step:rel-select}

\STATE\label{step:rel-reduce} Calculate the reduced ideal $\a_\v$ in the ideal class
$[\b]\cdot [\f^\v]$

\STATE Set $\Smooth \leftarrow \Smooth \union N(\a_\v)$

\ENDFOR\label{step:rel-endfor}

\STATE\label{step:rel-bernstein} Using Bernstein's algorithm~\cite{smoothparts}, find a
$\Primes$-smooth element $N(\a_\v) \in \Smooth$ (if one exists),
or else return $\nil$

\STATE\label{step:rel-factor} Find the prime factorization of the
integer $N(\a_\v)$

\STATE\label{step:rel-seysen} Using Theorem 3.1 of
Seysen~\cite{seysen} on the prime factorization of $N(\a_\v)$, factor
the ideal $\a_\v$ over $\f$ to obtain $\a_\v = \f^\aa$ for some $\aa
\in \Z^f$

\STATE Return $\z = \aa-\v$
\end{algorithmic}
\end{algorithm}

\paragraph{Computing $j(E')$.} Algorithm~\ref{alg:main} is the main
algorithm for evaluating the isogeny star operator. It takes as input
a discriminant $\Delta < 0$, an ideal class $[\b] \in \Cl(\O_\Delta)$,
and a $j$-invariant $j(E) \in \Ell_{q,n}(\O_\Delta)$, and produces the
element $j(E') \in \Ell_{q,n}(\O_\Delta)$ such that $[\b] * j(E) =
j(E')$. Eliminating the primes dividing $qn$ is necessary for the
computation of the isogenies in the final step of the algorithm.

Algorithm~\ref{alg:main} is correct since the ideals $\b$ and $\f^\z$
belong to the same ideal class, and thus act identically on
$\Ell_{q,n}(\O_\Delta)$.

\begin{algorithm}[t]
\caption{Computing $j(E')$}
\label{alg:main}
\begin{algorithmic}[1]
  \REQUIRE $\Delta$, $q$, $[\b]$, and a $j$-invariant $j(E) \in
  \Ell_{q,n}(\O_\Delta)$

\ENSURE The element $j(E') \in \Ell_{q,n}(\O_\Delta)$ such that
$[\b] * j(E) = j(E')$

\STATE Using Algorithm~\ref{alg:relation} with any valid choice of
$t$, compute a relation $\z \in \Z^f$ such that $[\b] = [\f^\z] =
[\p_1^{z_1} \p_2^{z_2} \cdots \p_f^{z_f}]$

\STATE\label{step:main-seq} Compute a sequence of isogenies $(\phi_1, \ldots, \phi_s)$ such
that the composition $\phi_c\colon E \to E_c$ of the sequence has
kernel $E[\p_1^{z_1} \p_2^{z_2} \cdots \p_f^{z_f}]$, using the method
of~\cite[\S 3]{BCL}

\STATE Return $j(E_c)$

\end{algorithmic}
\end{algorithm}

\subsection{Runtime analysis}\label{subsec:runtime}

Here we determine the theoretical running time of
Algorithm~\ref{alg:main}, as well as the optimal value of the
parameter $z$ in Algorithm~\ref{alg:relation}.  As is typical
for subexponential-time factorization algorithms involving a factor
base, these two quantities depend on each other, and hence both are
calculated simultaneously.

\begin{prop}\label{prop:relation-runtime}
The running time of Algorithm~\ref{alg:relation} is at most
$L(z) + L(\frac{1}{4z})$, assuming GRH.
\end{prop}
\begin{proof}
\step{factorbase} of Algorithm~\ref{alg:relation} takes
  time $L(z)$ \cite[Lemmas 11.3.1 and 11.3.2]{buchmann}.
  \step{rel-init1} of the algorithm requires $L(z)$ norm
  computations. \step{rel-init2} is negligible. \step{rel-reduce}
  requires $C \ln |\Delta|$ multiplications in the class group, each
  of which requires $O((\ln |\Delta|)^{1+\varepsilon})$ bit
  operations~\cite{schonhage}. Hence the \textbf{for} loop in
  \steprange{rel-startfor}{rel-endfor} has running time
  $L(\frac{1}{4z}) \cdot O((\ln |\Delta|)^{2+\varepsilon})$.
  Bernstein's algorithm~\cite{smoothparts} in \step{rel-bernstein} has
  a running time of $b(\log_2 b)^{2+\varepsilon}$ where $b = L(z) +
  L(\frac{1}{4z})$ is the combined size of $\Smooth$ and $\Primes$.
  Finding the prime factorization in \step{rel-factor} costs
  $L(z)$ using trial division, and Seysen's
  algorithm~\cite[Thm. 3.1]{seysen} in \step{rel-seysen} has
  negligible cost under ERH (and hence GRH).  Accordingly, we find
  that the running time is
\begin{align*}\textstyle
  L(z) + O((\ln |\Delta|)^{2+\varepsilon}) \cdot
  L(\frac{1}{4z}) + b(\log_2 b)^{2+\varepsilon} +
  L(z)
  = L(z) + L(\frac{1}{4z}),
\end{align*}
as desired.
\end{proof}

\begin{rem}\label{rem:relation}
  If we use quantum algorithms, then the performance boost obtained
  from Bernstein's algorithm is not necessary, since quantum computers
  can factor integers in polynomial time \cite{shor}. This allows for some
  simplification in Algorithm~\ref{alg:relation} in the quantum
  setting: there is no need to store elements of $\Smooth$ (since one
  can test directly for smooth integers via factoring),
  and the algorithm no longer requires superpolynomial space.
\end{rem}

\begin{prop}\label{prop:relation-success-probability}
  Under GRH, the probability that a single iteration of the
  \textbf{for} loop of Algorithm~\ref{alg:relation} produces an
  $\f$-smooth ideal $\a_\v$ is at least $L(-\frac{1}{4z})$.
\end{prop}
\begin{proof}
  We adopt the notation used in Theorem~\ref{thm:grh-graphs} and
  Corollary~\ref{cor:grh-graphs}. Apply Corollary~\ref{cor:grh-graphs}
  with the values $m = qn$, $B = 3$, and $x = f = L(z) \gg
  (\ln |\Delta|)^B$. Observe that $m$ has at most $O(\log q)$ prime
  divisors, and
\[
O(\log q) \ll L_q(\tfrac{1}{2},z(\tfrac{1}{2}-\varepsilon)) \le
L(z(\tfrac{1}{2}-\varepsilon)) = x^{1/2-\varepsilon}.
\]
Therefore Corollary~\ref{cor:grh-graphs} applies.  The
ideal class $[\b] \cdot [\f^\v]$ is equal to the ideal class obtained
by taking the walk of length $t$ in the Cayley graph
$\operatorname{Cay}(G, S_x)$, having initial vertex $[\b]$, and whose
edges correspond to the nonzero coordinates of the vector $\v$. Hence
a random choice of vector $\v$ under the constraints of
Algorithm~\ref{alg:relation} yields the same probability distribution
as a random walk in $\operatorname{Cay}(G, S_x)$ starting from $[\b]$.

  Let $S$ be the set of reduced ideals in $G$ with
  $L(z)$-smooth norm. By~\cite[Lemma
  11.4.4]{buchmann}, $
|S| \geq \sqrt{|\Delta|} L_{|\Delta|}(\frac12,-\frac{1}{4z}) \geq
\sqrt{|\Delta|} L(-\frac{1}{4z})
$.
Hence, by Corollary~\ref{cor:grh-graphs}, the probability that $\a_\v$ lies
in $S$ is at least
\[
\frac{1}{2} \frac{|S|}{|G|} \ge \frac{1}{2} \cdot
\frac{\sqrt{|\Delta|}}{|G|} \cdot L(-\tfrac{1}{4z}).
\]
Finally, Theorem~9.3.11 of~\cite{buchmann} states that
$\frac{\sqrt{|\Delta|}}{|G|} \ge
 \frac{1}{\ln|\Delta|}$. Hence the probability that $\a_\v$ is
$\f$-smooth is at least
\[
\frac{1}{2} \cdot \frac{1}{\ln|\Delta|} \cdot 
L(-\tfrac{1}{4z}) = 
L(-\tfrac{1}{4z}),
\]
as desired.
\end{proof}

\begin{cor}\label{cor:relation-success-probability}
  Under GRH, Algorithm~\ref{alg:relation}
  succeeds with probability at least $1 - \frac{1}{e}$.
\end{cor}
\begin{proof}
  Algorithm~\ref{alg:relation} loops through $\ell = L(\frac{1}{4z})$ 
  vectors $\v$, and by 
  Proposition~\ref{prop:relation-success-probability}, each such 
  choice of $\v$ has an independent
  $1/\ell$ chance of producing a smooth ideal $\a_\v$. Therefore the
  probability of success is at least
$
1 - \left(1 - \frac{1}{\ell}\right)^\ell > 1-\frac{1}{e}
$
as claimed.
\end{proof}

The following proposition shows that the relation vector $\z$ produced
by Algorithm~\ref{alg:relation} is guaranteed to have small coefficients.

\begin{prop}\label{prop:relation-norm}
Any vector $\z$ output by
Algorithm~\ref{alg:relation} satisfies $|\z|_1 <
(C+1) \ln |\Delta|$.
\end{prop}
\begin{proof}
Since $\z = \aa - \v$, we have $|\z|_1 \leq |\aa|_1 +
|\v|_1$.  But $|\v|_1 \leq C \ln |\Delta|$ by construction,
and the norm of $\a_\v$ is less than $\sqrt{|\Delta|/3}$~\cite[Prop.
9.1.7]{buchmann}, so
\[
|\aa|_1 < \log_2 \sqrt{|\Delta|/3} < \log_2 \sqrt{|\Delta|} <
\ln|\Delta|.
\]
This completes the proof.
\end{proof}

Finally, we analyze the running time of Algorithm~\ref{alg:main}.

\begin{theorem}\label{thm:main}
  Under GRH, Algorithm~\ref{alg:main} succeeds with probability at
  least $1-\frac{1}{e}$ and runs in time at most
\[ L(\tfrac{1}{4z}) +
\max\{L(3z), L(z) (\ln q)^{3+\varepsilon}\}.
\]
\end{theorem}

\begin{proof}
  We have shown that Algorithm~\ref{alg:relation} has running time
  $L(z) + L(\frac{1}{4z})$ and success
  probability at least $1-\frac{1}{e}$.
  Assuming that it succeeds, the computation of the individual
  isogenies $\phi_i$ in \step{main-seq} of Algorithm~\ref{alg:main}
  proceeds in one of two ways, depending on whether the characteristic
  of $\F_q$ is large~\cite[\S 3.1]{BCL}\cite[\S 3]{GHS} or
  small~\cite[\S 3.2]{BCL}. The large characteristic algorithm fails
  when the characteristic is small, whereas the small characteristic
  algorithm succeeds in all situations, but is slightly slower in
  large characteristic. For simplicity, we consider only the latter, and more
  general, algorithm.

  The general algorithm proceeds in two steps. In the first step, we
  compute the kernel polynomial of the isogeny. The time to perform
  one such calculation is $O((\ell
  (\ln q)\max(\ell, \ln q)^2)^{1+\varepsilon})$ in all cases
  (\cite[Thm. 1]{ls09}
  for characteristic $\geq 5$ and~\cite[Thm. 1]{defeo} for
  characteristic $2$ or $3$).  In the second step, we compute the
  equation of the isogenous curve using V\'elu's
  formulae~\cite{velu}. This second step has a
  running time of $O(\ell^{2+\varepsilon} (\ln
  q)^{1+\varepsilon})$~\cite[p.  214]{pairingvolcano}. Hence the
  running time of \step{main-seq} is at most
\[
|\z|_1 (O((\ell (\ln q)\max(\ell, \ln q)^2)^{1+\varepsilon}) +
O(\ell^{2+\varepsilon} (\ln q)^{1+\varepsilon})).
\]
By Proposition~\ref{prop:relation-norm}, this expression is at most
\begin{align*}
&(C+1)
(\ln|\Delta|) (\max\{L(3z), L(z) (\ln
q)^{3+\varepsilon}\} + L(2z) (\ln q)^{1+\varepsilon}) \\
&=\max\{L(3z), L(z) (\ln
q)^{3+\varepsilon}\}.
\end{align*}
The theorem follows.
\end{proof}

\begin{cor}\label{cor:main}
  Under GRH, Algorithm~\ref{alg:main} has a worst-case running time of
  at most $L_q(\frac{1}{2},\frac{\sqrt{3}}{2})$.
\end{cor}
\begin{proof}
Using the inequality $|\Delta| \leq 4q$, we may rewrite
Theorem~\ref{thm:main} in terms of $q$. We obtain the following upper
bound for the running time:
\[L(\tfrac{1}{4z}) +
\max\{L(3z), L(z) (\ln q)^{3+\varepsilon}\} \leq
L_q(\tfrac{1}{2},\tfrac{1}{4z} + 3z).
\]
The optimal choice of $z = \frac{1}{2\sqrt{3}}$ yields the running
time bound of $L_q(\frac{1}{2},\frac{\sqrt{3}}{2})$.
\end{proof}

\begin{rem}\label{rem:action-improvement}
  Using our technique for eliminating heuristics,
  Bisson~\cite{bisson-2011} has recently developed a
  subexponential-time algorithm for determining endomorphism rings of
  elliptic curves, assuming only GRH. As part of that work, Bisson
  presents a faster algorithm~\cite[Prop. 4.4]{bisson-2011} for
  determining the curves appearing in the sequence of isogenies in
  Step~\ref{step:main-seq} of Algorithm~\ref{alg:main}, with running
  time quadratic in the isogeny degrees, improving upon the cubic
  time required in prior algorithms. Using this
  algorithm, the running time of Algorithm~\ref{alg:main} improves to
  $L_q(\frac{1}{2},\frac{1}{\sqrt{2}})$.
\end{rem}

\begin{rem}\label{rem:evaluate}
  Our algorithm for computing the isogeny star operator readily
  extends to an algorithm for evaluating isogenies in subexponential
  time. As in~\cite{BCL,jv}, we specify an isogeny
    $\phi\colon E \to E'$ by providing the ideal $\b \subset \End(E) =
    \O_\Delta$ corresponding to the kernel of $\phi$.  To distinguish
    between isogenies that are identical up to isomorphism, we define
    a \emph{normalized} isogeny~\cite{bostan,BCL} to be one where
    $\phi^*(w_{E'}) = w_E$.  Algorithm~\ref{alg:main} applied to the
    input $\b$ yields an (unnormalized) isogeny $\phi_c\colon E \to
    E_c$ isomorphic to the desired isogeny $\phi$. To find the
    normalized isogeny, we must evaluate the necessary isomorphism
    explicitly. This can be easily done by using~\cite[Algorithm 3,
    Steps 20--23]{jv} in conjunction with~\cite[Algorithm 4.1, Steps
    4--6]{BCL} on the relation produced by
  Algorithm~\ref{alg:relation}. These additional steps are not
  rate-limiting, so the running time of the algorithm is
  unchanged. Bisson's improvement
  (Remark~\ref{rem:action-improvement}) does not apply here, since we
  need to evaluate the actual isogeny, rather than just find the
  isogenous curve.
\end{rem}

\section{A quantum algorithm for constructing
  isogenies}\label{sec:quantum}

Our quantum algorithm for constructing isogenies uses a simple reduction to the
abelian hidden shift problem.  To define this problem, let $A$ be a
known finite abelian group (with the group operation written multiplicatively) and let $f_0,f_1\colon  A \to S$ be black-box
functions, where $S$ is a known finite set.  We say that $f_0,f_1$
hide a shift $s \in A$ if $f_0$ is injective and $f_1(x)=f_0(xs)$
(i.e., $f_1$ is a shifted version of $f_0$).  The goal of the hidden
shift problem is to determine $s$ using queries to such black-box
functions.  Note that this problem is equivalent to the hidden subgroup problem in the $A$-dihedral group, the nonabelian group $A \rtimes \Z_2$ where $\Z_2$ acts on $A$ by inversion.

Isogeny construction is easily reduced to the hidden shift problem
using the group action defined in \sec{background}.  Given horizontally isogenous
curves $E_0,E_1$ with endomorphism ring $\O_\Delta$, we define
functions $f_0,f_1\colon \Cl(\O_\Delta) \to \Ell_{q,n}(\O_\Delta)$ that hide
$[\mathfrak{s}] \in \Cl(\O_\Delta)$, where $[\mathfrak{s}]$ is the
ideal class such that $[\mathfrak{s}] * j(E_0) = j(E_1)$.
Specifically, let $f_c([\b]) = [\b] * j(E_c)$.  Then it is immediate
that $f_0,f_1$ hide $[\mathfrak{s}]$:

\begin{lemma}\label{lem:hiding}
  The function $f_0$ is injective and $f_1([\b]) =
  f_0([\b][\mathfrak{s}])$.
\end{lemma}

\begin{proof}
Since $*$ is a group action,
\begin{align*}
  f_1([\b])
  &= [\b] * j(E_1) \\
  &= [\b] * ([\mathfrak{s}] * j(E_0)) \\
  &= ([\b][\mathfrak{s}]) * j(E_0) \\
  &= f_0([\b][\mathfrak{s}]).
\end{align*}
If there are distinct ideal classes $[\b],[\b']$ such that
$f_0([\b])=f_0([\b'])$, then $[\b] * j(E_0) = [\b'] * j(E_0)$, which
contradicts the fact that the action is free and transitive~\cite[Thm.
4.5]{waterhouse}.  Thus $f_0$ is injective.
\end{proof}

Note that a similar connection between isogenies and hidden shift
problems was described in \cite[Section 7.2]{stolbunov}.  However,
that paper did not recognize the significance of the reduction, and in
particular did not appreciate the role played by injectivity.  Without
the assumption that $f_0$ is injective, the hidden shift problem can
be as hard as the search problem, and hence requires exponentially
many queries \cite{bbbv} (although for non-injective
  functions $f_0$ with appropriate structure, such as the Legendre
  symbol, the non-injective hidden shift problem can be solved by a
  quantum computer in polynomial time \cite{DHI03}).  On the other
hand, injectivity implies that the problem has polynomial quantum
query complexity \cite{EH00}, allowing for the possibility of faster
quantum algorithms.

This reduction allows us to apply quantum algorithms for the hidden
shift problem to construct isogenies. The (injective) hidden shift problem
can be solved in quantum subexponential time assuming we can
evaluate the group action in subexponential time.  The latter is possible
due to \alg{main}.

We consider two different approaches to solving the hidden shift
problem in subexponential time on a quantum computer. The first, due
to Kuperberg~\cite{kuperberg}, has a faster running time but requires
superpolynomial space. The second approach generalizes an algorithm of
Regev~\cite{regev-polyspace}. It uses only polynomial space, but is
slower than Kuperberg's original algorithm.

\paragraph{Method 1: Kuperberg's algorithm.} 
Kuperberg's approach to the abelian hidden shift problem is based on the idea of performing a
Clebsch-Gordan sieve on coset states. The following appears as Theorem 7.1 of~\cite{kuperberg}.
\begin{theorem}\label{thm:hiddenshift-kuperberg}
  The abelian hidden shift problem has a [quantum] algorithm with time
  and query complexity $2^{O(\sqrt{n})}$, where $n$ is the length of
  the output, uniformly for all finitely generated abelian groups.
\end{theorem}
 In our context, $2^{O(\sqrt{n})} = 2^{O(\sqrt{\ln
      |\Delta|})}$ since $|{\Cl(\O_\Delta)}| = O(\sqrt\Delta
  \ln\Delta)$ \cite[Theorem~9.3.11]{buchmann}.  Furthermore,
  $2^{O(\sqrt{\ln |\Delta|})} = L(o(1)) = L(0)$ regardless of the
  value of the implied constant in the exponent, since the exponent on
  the left has no $\sqrt{\ln \ln |\Delta|}$ term, whereas $L(0)$ does.
  As mentioned above, Kuperberg's algorithm also requires
  superpolynomial space (specifically, it uses $2^{O(\sqrt{n})}$
  qubits).

\paragraph{Method 2: Regev's algorithm.}  Regev~\cite{regev-polyspace}
showed that a variant of Kuperberg's sieve leads to a slightly slower
algorithm using only polynomial space.  In particular, he proved
\thm{hiddenshift} below in the case where $A$ is a cyclic group whose order
is a power of $2$ (without giving an explicit value for
the constant in the exponent).
\thm{hiddenshift} generalizes Regev's
algorithm to arbitrary finite abelian groups. A detailed proof of
\thm{hiddenshift} appears 
in the Appendix (see \thm{hiddenshift-alg}).

\begin{theorem}\label{thm:hiddenshift}
  Let $A$ be a finite abelian group and let functions $f_0,f_1$ hide
  some unknown $s \in A$.  Then there is a quantum algorithm that
  finds $s$ with time and query complexity $L_{|A|}(\frac12,\sqrt2)$ 
  using space $\poly(\log|A|)$.
\end{theorem}

We now return to the original problem of constructing isogenies. Note
that to use the hidden shift approach, the group structure of
$\Cl(\O_\Delta)$ must be known. Given $\Delta$, it is straightforward
to compute $\Cl(\O_\Delta)$ using existing quantum algorithms
(see the proof of Theorem~\ref{thm:quantum}). Thus, we assume for
simplicity that the discriminant $\Delta$ is given as part of the
input. This requirement poses no difficulty, since all existing
proposals for isogeny-based public-key
cryptosystems~\cite{hhs,rost,stolbunov} stipulate that $\O_\Delta$ is
a maximal order, in which case its discriminant can be computed
easily: simply calculate the trace $t(E)$ of the curve using Schoof's
algorithm \cite{schoof}, and factor $t(E)^2-4q$ to obtain the
fundamental discriminant $\Delta$ (note of course that factoring is
easy on a quantum computer \cite{shor}).

\begin{rem}
  One can conceivably imagine a situation where one is asked to
  construct an isogeny between two given isogenous curves of unknown
  but identical endomorphism ring. Although we are not aware of any
  cryptographic applications of this scenario, it presents no
  essential difficulty. 
  Bisson~\cite{bisson-2011} has shown using \corol{grh-graphs} that
  the discriminant $\Delta$ of an elliptic curve can be computed in
  $L_q(\frac{1}{2},\frac{1}{\sqrt2})$ time under only GRH (assuming
  that factoring is easy).
\end{rem}

Assuming $\Delta$ is known, we decompose $\Cl(\O_\Delta)$ as a direct
sum of cyclic groups, with a known generator for each, and then solve 
the hidden shift problem.  The overall
procedure is described in \alg{quantum}.

\begin{algorithm}[t]
  \caption{Isogeny construction}
  \label{alg:quantum}
  \begin{algorithmic}[1]
  \REQUIRE
  A finite field $\F_q$,
  a discriminant $\Delta < 0$, and
  Weierstrass equations of horizontally isogenous elliptic curves $E_0,E_1$
  
  \ENSURE
  $[\mathfrak{s}] \in \Cl(\O_\Delta)$ such that $[\mathfrak{s}] *
  j(E_0) = j(E_1)$
  
  \STATE \label{step:classgroup} Decompose $\Cl(\O_\Delta) =
  \langle[\b_1]\rangle \oplus \cdots \oplus \langle[\b_k]\rangle$
  where $|\<[\b_j]\>| = n_j$

  \STATE \label{step:hiddenshift} Solve the hidden shift problem
  defined by functions $f_0,f_1\colon \Z_{n_1} \times \cdots \times \Z_{n_k}
  \to \Ell_{q,n}(\O_\Delta)$ satisfying $f_c(x_1,\ldots,x_k) =
  ([\b_1]^{x_1} \cdots [\b_k]^{x_k}) * j(E_c)$, giving some
  $(s_1,\ldots,s_k) \in \Z_{n_1} \times \cdots \times \Z_{n_k}$
  
  \STATE \label{step:output} Output $[\mathfrak{s}] =
  [\mathfrak{b}_1]^{s_1} \cdots [\mathfrak{b}_k]^{s_k}$
\end{algorithmic}
\end{algorithm}

\begin{theorem}\label{thm:quantum}
  Assuming GRH, \alg{quantum} runs in time
  $L_q(\frac12,\frac{\sqrt3}{2})$ (respectively,
  $L_q(\frac12,\frac{\sqrt3}{2}+\sqrt2)$) using \thm{hiddenshift-kuperberg}
  (respectively, \thm{hiddenshift}) to solve the hidden shift problem.
\end{theorem}

\begin{proof}
  We perform \step{classgroup} using \cite[Algorithm 10]{mosca}, which
  determines the structure of an abelian group given a generating set
  and a unique representation for the group elements.  We represent
  the elements uniquely using reduced quadratic forms, and we use the
  fact that, under ERH (and hence GRH), the set of ideal classes of
  norm at most $6 \ln^2 |\Delta|$ forms a generating set \cite[p.~376]{bach}.
  By \thm{hiddenshift-kuperberg} (resp.\ \thm{hiddenshift}),
  \step{hiddenshift} uses $L(o(1)) = L(0)$ (resp.\ $L(\sqrt2)$)
  evaluations of the functions $f_i$.  By \corol{main},
  these functions can be evaluated in time
  $L_q(\frac12,\frac{\sqrt3}{2})$ using \alg{main}, assuming GRH.
  Overall, \step{hiddenshift} takes time
  $L_q(\frac12,\frac{\sqrt3}{2}+o(1)) = L_q(\frac12,\frac{\sqrt3}{2})$ if
  \thm{hiddenshift-kuperberg} is used, or
  $L_q(\frac12,\frac{\sqrt3}{2}+\sqrt2)$ if \thm{hiddenshift} is used.  The
  cost of \step{output} is negligible.
\end{proof}

\begin{rem}
  Using the improved algorithm for evaluating the isogeny star operator described in \rmrk{action-improvement}, the running time of \alg{quantum} is improved to $L_q(\frac12,\frac{1}{\sqrt2} + o(1)) = L_q(\frac12,\frac{1}{\sqrt2})$ using \thm{hiddenshift-kuperberg} to solve the hidden shift problem (requiring superpolynomial space), and to $L_q(\frac12,\frac{1}{\sqrt2}+\sqrt2) = L_q(\frac12,\frac{3}{\sqrt 2})$ using \thm{hiddenshift} (requiring only polynomial space).
\end{rem}

\begin{rem}\label{rem:improvements}
  The running time of the algorithm is ultimately limited by
  two factors: the best known quantum algorithm for the hidden shift
  problem runs in superpolynomial time, and the same holds for the best
  known (classical or quantum) algorithm for computing the isogeny star 
  operator. Improving only one of these results to
  take polynomial time would still result in a superpolynomial-time
  algorithm.
\end{rem}

\section*{Acknowledgments}

This work was supported in part by MITACS, NSERC, the Ontario 
Ministry of Research and Innovation, QuantumWorks, and the US 
ARO/DTO.

\appendix

\section{Subexponential-time and polynomial-space quantum algorithm for the general abelian hidden shift problem}
\label{app:polyspace}

Following Kuperberg's discovery of a subexponential-time quantum algorithm for the hidden shift problem in any finite abelian group $A$ \cite{kuperberg}, Regev presented a modification of Kuperberg's algorithm that requires only polynomial space, with a slight increase in the running time \cite{regev-polyspace}.  However, Regev only explicitly considered the case $A=\Z_{2^n}$, and while he showed that the running time is $L_{|A|}(\frac12,c)$, he did not determine the value of the constant $c$.

In this appendix we describe a polynomial-space quantum algorithm for the general abelian hidden shift problem using time $L_{|A|}(\frac12,\sqrt2)$.
We use several of the same techniques employed by Kuperberg \cite[Algorithm~5.1 and Theorem~7.1]{kuperberg} to go beyond the case $A=\Z_{2^n}$, adapted to work with a Regev-style sieve that only uses polynomial space.

Let $A = \Z_{N_1} \times \cdots \times \Z_{N_t}$ be a finite abelian group.  Consider the hidden shift problem with hidden shift $s = (s_1,\ldots,s_t) \in A$.  By Fourier sampling, one (coherent) evaluation of the hiding functions $f_0,f_1$ can produce the state
\begin{align}\label{eq:hsqubit}
  |\psi_x\> \defeq \frac{1}{\sqrt2} \left(|0\> + \exp\left[2\pi i\left(\frac{s_1 x_1}{N_1}+\cdots+\frac{s_t x_t}{N_t}\right)\right]|1\>\right)
\tag{$\psi$}
\end{align}
with a known value $x = (x_1,\ldots,x_t) \inrand A$ (see for example the proof of Theorem 7.1 in \cite{kuperberg}), where $x \inrand A$ denotes that $x$ occurs uniformly at random from $A$.
For simplicity, we begin by considering the case where $A = \Z_N$ is cyclic.  Then Fourier sampling produces states
\begin{align*}
  |\psi_x\> = \frac{1}{\sqrt2}(|0\> + \omega^{sx} |1\>)
\end{align*}
where $x \inrand \Z_N$ is known and $\omega \defeq e^{2\pi i/N}$.

If we could make states $|\psi_x\>$ with chosen values of $x$, then we could determine $s$.  In particular, the following observation is attributed to Peter H{\o}yer in \cite{kuperberg}:

\begin{lemma}\label{lem:qftreconstruct}
Given one copy each of the states $|\psi_1\>,|\psi_2\>,|\psi_4\>,\ldots,|\psi_{2^{k-1}}\>$, where 
$2^k = \Omega(N)$, 
one can reconstruct $s$ in polynomial time with probability 
$\Omega(1)$.
\end{lemma}

\begin{proof}
We have
\begin{align*}
  \bigotimes_{j=0}^{k-1} |\psi_{2^j}\>
    &= \frac{1}{\sqrt{2^k}} \sum_{y=0}^{2^k-1} \omega^{sy} |y\>.
\end{align*}
Apply the inverse quantum Fourier transform over $\Z_N$ (which runs in
$\poly(\log N)$ time \cite{kitaev}) and measure in the computational basis.
The Fourier transform of $|s\>$, namely $\frac{1}{\sqrt{N}}\sum_{y=0}^{N-1} \omega^{sy}|s\>$, has overlap squared with this state of $2^k/N$, which implies the claim.
\end{proof}

We aim to produce states of the form $|\psi_{2^j}\>$ using a sieve that combines states to prepare new ones with more desirable labels.
A basic building block is \alg{dcomb}, which can be used to produce states with smaller labels.

\begin{algorithm}[t]
\caption{Combining states to give smaller labels}
\label{alg:dcomb}
\begin{algorithmic}[1]
  \REQUIRE Parameters $B,B'$ and states $|\psi_{x_1}\>,\ldots,|\psi_{x_k}\>$ with known $x_1,\ldots,x_k \inrand \{0,1,\ldots,B-1\}$
  \ENSURE State $|\psi_{x'}\>$ with known $x' \inrand \{0,1,\ldots,B'-1\}$
  \IF{$\exists i\colon x_i \ge 2B' \floor{B/2B'}$}
    \STATE\label{step:dcomb-unif} Abort
  \ENDIF
  \STATE Introduce an ancilla register and compute
  \[
    \frac{1}{\sqrt{2^k}} \sum_{y \in \{0,1\}^k} \omega^{s (x \cdot y)} |y\>|\floor{(x \cdot y)/2B'}\>
  \]
  where $x \cdot y \defeq \sum_{i=1}^k x_i y_i$
  \STATE Measure the ancilla register, giving an outcome $q$ and a state
  \[
    \frac{1}{\sqrt \nu} \sum_{j=1}^\nu \omega^{s (x \cdot y^j)} |y^j\>
  \]
  where $y^1,\ldots,y^\nu \ne 0^k$ are the $k$-bit strings such that $\floor{(x \cdot y^j)/2B'}=q$
  \STATE\label{step:dcomb-subsetsum} Compute $y^1,\ldots,y^\nu$ by brute force
  \IF{$\nu=1$}
    \STATE\label{step:dcomb-onesol} Abort
  \ENDIF
  \STATE\label{step:dcomb-project} Project onto $\spn\{|y^1\>,|y^2\>\}$ or $\spn\{|y^3\>,|y^4\>\}$ or \ldots\ or $\spn\{|y^{2\floor{\nu/2}-1}\>,|y^{2\floor{\nu/2}}\>\}$, giving an outcome $\spn\{|y^\star\>,|y^\varstar\>\}$
  \STATE\label{step:dcomb-outputlabel} Let $x' = x \cdot (y^\varstar - y^\star)$ where $x \cdot y^\varstar \ge x \cdot y^\star$ WLOG
  \IF{$x' \in \{1,\ldots,B'-1\}$}\label{step:dcomb-poststart}
    \STATE Abort with probability $B'/(2B'-x')$
  \ELSIF{$x' \in \{B',\ldots,2B'-1\}$}
    \STATE Abort
  \ENDIF\label{step:dcomb-postend}
  \STATE\label{step:dcomb-final} Relabel $|y^\star\> \mapsto |0\>$ and $|y^\varstar\> \mapsto |1\>$, giving a state $|\psi_{x'}\>$
\end{algorithmic}
\end{algorithm}

\begin{lemma}\label{lem:dcomb}
\alg{dcomb} runs in time $2^k \poly(\log N)$ and succeeds with probability $\Omega(1)$ provided $4k \le B/B' \le 2^k/k$.
\end{lemma}

\begin{proof}
The running time is dominated by the brute force calculation in \step{dcomb-subsetsum} and the projection in \step{dcomb-project}, both of which can be performed in time $2^k \poly(\log N)$.

The probability of aborting in \step{dcomb-unif} for any one $x_i$ is $1-\frac{2B'}{B}\floor{\frac{B}{2B'}} \le \frac{2B'}{B}$, so by the union bound, the overall probability of aborting in this step is at most $k\frac{2B'}{B} \le 1/2$.
Conditioned on not aborting in \step{dcomb-unif}, $x_i \inrand \{0,1,\ldots,2B'\floor{B/2B'}-1\}$.

Let $x \cdot y^j = q (2B') + r^j$ where $0 \le r^j < 2B'$ ($q$ is the measurement outcome, which is independent of $j$).  By the uniformity of the $x_i$s, each $r^j = x \cdot y^j \bmod 2B'$ is uniformly distributed over $\{0,1,\ldots,2B'-1\}$.  Thus the output label is $x' = x \cdot(y^\varstar - y^\star) = |r^\varstar-r^\star|$ where $r^\star,r^\varstar \inrand \{0,1,\ldots,2B'-1\}$.  A simple calculation shows that the distribution of $|r^\varstar - r^\star|$ is
\[
  \Pr(|r^\varstar - r^\star| = \Delta) = \begin{cases}
  \tfrac{1}{2B'} & \text{for $\Delta=0$} \\
  \tfrac{2B'-\Delta}{2B'^2} & \text{for $\Delta \in \{1,\ldots,2B'-1\}$}.
  \end{cases}
\]
Thus the probability that we abort in \steprange{dcomb-poststart}{dcomb-postend} is $1/2$, and conditioned on not aborting in these steps, $x' \inrand \{0,1,\ldots,B'-1\}$.  Thus the algorithm is correct if it reaches \step{dcomb-final}.

It remains to show that the algorithm succeeds with constant probability.  We have already bounded the probability that we abort in \step{dcomb-unif} and \steprange{dcomb-poststart}{dcomb-postend}.
Since $y=0$ occurs with probability $2^{-k}$ and at most one state $|y^\nu\>$ can be unpaired (and this only happens when $\nu$ is odd), the projection in \step{dcomb-project} fails with probability at most $\nu^{-1}+2^{-k} \le 1/3 + o(1)$.
We claim that the  probability of aborting in \step{dcomb-onesol} (i.e., the probability that $\nu=1$) is also bounded away from $1$.
Call a value of $q$ bad if $\nu=1$.  Since $0 \le x \cdot y \le k(B-1)$, there are at most $kB/2B'$ possible values of $q$, and in particular, there can be at most $kB/2B'$ bad values of $q$.  Since the probability of any particular bad $q$ is $1/2^k$, the probability that $q$ is bad is at most $k B/B'2^{k+1} \le 1/2$.  This completes the proof.
\end{proof}

We apply this combination procedure using the generalized sieve of \alg{sieve}, which is equivalent to Regev's ``pipeline of routines'' \cite{regev-polyspace}.

\begin{algorithm}[t]
\caption{Sieving quantum states}
\label{alg:sieve}
\begin{algorithmic}[1]
  \REQUIRE Procedures to prepare states
  from a set $S_0$ and to combine $k$ states
  from $S_{i-1}$ to make a state 
  from $S_i$ with probability at least $p$ for each $i \in \{1,\ldots,m\}$
  \ENSURE State from $S_m$
  \REPEAT
    \WHILE{for all $i$ we have fewer than $k$ states from $S_i$}
      \STATE Make a state from $S_0$
    \ENDWHILE
    \STATE Combine $k$ states from some $S_i$ to make a state from $S_{i+1}$ with probability at least $p$
  \UNTIL{there is a state from $S_m$}
\end{algorithmic}
\end{algorithm}

\begin{lemma}\label{lem:sieve}
Suppose $m e^{-2k} = o(1)$.
Then \alg{sieve} is correct, succeeds with probability $1-o(1)$ using $k^{(1+o(1))m}$ state preparations and combination operations, and uses space $O(mk)$.
\end{lemma}

\begin{proof}
  If \alg{sieve} outputs a state from $S_m$ then it is correct.
  Since the algorithm never stores more than $O(mk)$ states at a time, it uses space $O(mk)$.
  It remains to show that the algorithm is likely to succeed using only $k^{(1+o(1))m}$ state preparations and combination operations.
  
  If we could perform combinations deterministically, we would need
\begin{align*}
  \text{$1$ state~} &\text{from $S_m$,}\\
  \text{$k$ states~} &\text{from $S_{m-1}$,}\\
  \text{$k^2$ states~} &\text{from $S_{m-2}$,}\\
  &\vdots \\
  \text{$k^m$ states~} &\text{from $S_0$.}
\end{align*}
Since the combinations only succeed with probability $p$, we lower bound the probability of eventually producing $(2k/p)^{m-i}$ states from $S_i$ for each $i \in \{1,\ldots,m\}$ (so in particular, we produce one state from $S_m$).  Given $(2k/p)^{m-i+1}$ states from $S_{i-1}$, the expected number of successful combinations is $p(2k/p)^{m-i+1}/k = 2(2k/p)^{m-i}$, whereas only $(2k/p)^{m-i}$ successful combinations are needed.  By the Chernoff bound,
the probability of having fewer than $(2k/p)^{m-i}$ successful combinations is at most 
$e^{-p(2k/p)^{m-i}}$.  Thus, by the union bound, the probability that the algorithm fails is at most
\begin{align*}
  \sum_{i=1}^{m-1} e^{-p(2k/p)^{m-i}}
  &\le m e^{-2k},
\end{align*}
so the probability of success is $1-o(1)$.

Finally, the number of states from $S_0$ is
$(2k/p)^m
= k^{(1+o(1))m}$
and the total number of combinations is $\sum_{i=0}^{m-1} (2k/p)^{m-i}/k = k^{(1+o(1))m}$.
\end{proof}

When using the sieve, we have the freedom to choose the relationship between $k$ and $m$ to optimize the running time.  Suppose that $mk = (1+o(1))\log_2 N$ (intuitively, to cancel $\log_2 N$ bits of the label), and also suppose that the combination operation takes time $2^k \poly(\log N)$ (as in \lem{dcomb}).  Then if we take $k = c \sqrt{\log_2 N \log_2\log_2 N}$, we find that the overall running time of \alg{sieve} is
$2^k 2^{(1+o(1)) m \log_2 k} \poly(\log N) = L_N(\tfrac12,c + \tfrac1{2c})$.
Choosing $c=\frac{1}{\sqrt2}$ gives the best running time, $L_N(\tfrac12,\sqrt2)$.

We now consider how to apply the sieve.  To use \lem{qftreconstruct}, our goal is to prepare states of the form $|\psi_{2^j}\>$ for each $j \in \{0,1,\ldots,\floor{\log_2 N}\}$.  First we show how to prepare the state $|\psi_1\>$ in time $L_N(\frac12,\sqrt2)$ using 
\alg{dcomb} as the combination procedure
in \alg{sieve}.  For $i \in \{0,1,\ldots,m\}$, the $i$th stage of the sieve produces states with labels from $S_i = \{0,1,\ldots,B_i-1\}$.  \lem{sievesizesd} below shows that there is a choice of the $B_i$ with $B_0=N$, $B_m=2$, and successive ratios of the $B_i$s satisfying the conditions of \lem{dcomb}, such that $2^k k^{(1+o(1))m} = L_N(\frac12,\sqrt2)$.  It then follows that \alg{sieve} produces a uniformly random label from $S_m=\{0,1\}$ with constant probability in time $L_N(\frac12,\sqrt2)$, and in particular, can be used to produce a copy of $|\psi_1\>$ in time $L_N(\frac12,\sqrt2)$.

\begin{lemma}\label{lem:sievesizesd}
There is a constant $N_0$ such that for all $N>N_0$, letting $B_i = \floor{N /\rho^i}$ where $\rho = (N/2)^{1/m}$ and
\begin{align*}
k &= \Floor{\sqrt{\tfrac{1}{2} \log_2 N \log_2\log_2 N}} &
m &= \Ceil{\frac{\log_2 N/2}{k-\log_2 2k}}
   = \Theta\left(\sqrt{\frac{\log_2 N}{\log_2\log_2 N}}\right),
\end{align*}
we have $B_0=N$, $B_m=2$, and $4k \le B_{i-1}/B_i \le 2^k/k$ for all $i\in \{1,\ldots,m\}$.
\end{lemma}

\begin{proof}
Clearly $B_0=N$, and the value of $\rho$ is chosen so that $B_m = 2$.

For $i \in \{1,\ldots,m\}$, we have
\begin{align*}
  \frac{B_{i-1}}{B_i}
  &= \frac{\floor{N/\rho^{i-1}}}{\floor{N/\rho^i}}
  \le \frac{N/\rho^{i-1}}{N/\rho^i - 1}
  = \frac{\rho}{1-\rho^i/N}.
\end{align*}
Since $\rho^i/N \le \rho^m/N = 1/2$, we have $B_{i-1}/B_i \le 2\rho$.  Then using
\begin{align*}
  \rho
  \le (N/2)^{\frac{k-\log_2 2k}{\log_2 N/2}}
  = \frac{2^k}{2k}
\end{align*}
gives $B_{i-1}/B_i \le 2^k/k$ as claimed.

Similarly, we have
\begin{align*}
  \frac{B_{i-1}}{B_i}
  = \frac{\floor{N/\rho^{i-1}}}{\floor{N/\rho^i}}
  \ge \frac{N/\rho^{i-1} - 1}{N/\rho^i}
  = \rho - \rho^i/N
  \ge \rho - \tfrac12.
\end{align*}
Since
\begin{align*}
  \rho
  = (N/2)^{\Theta(\sqrt{\log_2 \log_2 N / \log_2 N})}
  = 2^{\Theta(\sqrt{\log_2 N \log_2 \log_2 N})}
  = 2^{\Theta(k)},
\end{align*}
we have $\rho - \tfrac12 \ge 4k$ for sufficiently large $N$.
This completes the proof.
\end{proof}

If $N$ is odd, then division by $2$ is an automorphism of $\Z_N$.  Thus we can prepare $|\psi_{2^j}\>$ by performing the above sieve under the automorphism $x \mapsto 2^{-j} x$.  It follows that the abelian hidden shift problem in a cyclic group of odd order $N$ can be solved in time $L_N(\tfrac12,\sqrt2)$.

Now suppose that $N=2^n$ is a power of $2$.  In this case, we first use a combination procedure that zeros out low-order bits, as described in \alg{lsbcomb}.  We use the notation $xS \defeq \{xz: z \in S\}$ for any $x \in \Z$ and $S \subset \Z$.

\begin{algorithm}[t]
\caption{Combining states to cancel low-order bits}
\label{alg:lsbcomb}
\begin{algorithmic}[1]
  \REQUIRE Parameters $\ell,\ell'$ and states $|\psi_{x_1}\>,\ldots,|\psi_{x_k}\>$ with known $x_1,\ldots,x_k \inrand 2^\ell \{0,1,\ldots,N/2^\ell - 1\}$
  \ENSURE State $|\psi_{x'}\>$ with known $x' \inrand 2^{\ell'}\{0,1,\ldots,N/2^{\ell'}-1\}$
  \STATE Introduce an ancilla register and compute
  \[
    \frac{1}{\sqrt{2^k}} \sum_{y \in \{0,1\}^k} \omega^{s (x \cdot y)} |y\>|x \cdot y \bmod 2^{\ell'}\>
  \]
  \STATE Measure the ancilla register, giving an outcome $r$ and a state
  \[
    \frac{1}{\sqrt \nu} \sum_{j=1}^\nu \omega^{s (x \cdot y^j)} |y^j\>
  \]
  where $y^1,\ldots,y^\nu \ne 0^k$ are the $k$-bit strings such that $x \cdot y^j \bmod 2^{\ell'}=r$
  \STATE\label{step:lsbcomb-subsetsum} Compute $y^1,\ldots,y^\nu$ by brute force
  \IF{$\nu=1$}
    \STATE Abort
  \ENDIF
  \STATE\label{step:lsbcomb-project} Project onto $\spn\{|y^1\>,|y^2\>\}$ or $\spn\{|y^3\>,|y^4\>\}$ or \ldots\ or $\spn\{|y^{2\floor{\nu/2}-1}\>,|y^{2\floor{\nu/2}}\>\}$, giving an outcome $\spn\{|y^\star\>,|y^\varstar\>\}$
  \STATE\label{step:lsbcomb-final} Relabel $|y^\star\> \mapsto |0\>$ and $|y^\varstar\> \mapsto |1\>$, giving a state $|\psi_{x'}\>$ with $x' = x \cdot (y^\varstar-y^\star) \bmod N$
\end{algorithmic}
\end{algorithm}

\begin{lemma}\label{lem:lsbcomb}
\alg{lsbcomb} runs in time $2^k \poly(\log N)$ and succeeds with probability $\Omega(1)$ provided $k \ge \ell' - \ell + 1$.
\end{lemma}

\begin{proof}
The proof is similar to that of \lem{dcomb}.  Again the running time is dominated by the brute force calculation in \step{lsbcomb-subsetsum} and the projection in \step{lsbcomb-project}, both of which can be performed in time $2^k \poly(\log N)$.

We claim that the algorithm is correct if it reaches \step{lsbcomb-final}.
Observe that $x \cdot y^j \bmod N = q^j 2^{\ell'} + r$ where $r$ is independent of $j$.  Since $y^j \ne 0^k$, $x \cdot y^j \bmod N \inrand 2^\ell\{0,1,\ldots,N/2^\ell-1\}$, so $q^j \inrand \{0,1,\ldots,N/2^{\ell'}-1\}$, and hence $x' = (q^\varstar-q^\star)2^{\ell'} \bmod N \inrand 2^{\ell'}\{0,1,\ldots,N/2^{\ell'}-1\}$ as required.

The projection in \step{lsbcomb-project} fails with probability at most $1/3 + o(1)$.  It remains to show that the algorithm reaches \step{lsbcomb-project} with probability $\Omega(1)$, i.e., to upper bound the probability that $\nu=1$.  Call a value of $r$ bad if $\nu=1$.  There are $2^{\ell'-\ell}$ possible values of $r$, so in particular there are at most $2^{\ell'-\ell}$ bad values of $r$.  Since the probability of any particular bad $r$ is $1/2^k$, the probability that $r$ is bad is at most $2^{\ell'-\ell-k} \le 1/2$.  This completes the proof.
\end{proof}

\alg{lsbcomb} is similar to the combination procedure used in \cite{regev-polyspace}, but differs in that the latter requires $\nu=O(1)$, which is established in the analysis using a second moment argument.  The modification of pairing as many values of $y$ as possible allows us to use a simpler analysis (with essentially the same performance).

To produce a state of the form $|\psi_{2^j}\>$, we first use 
\alg{lsbcomb}
to cancel low-order bits and then use
\alg{dcomb}
to cancel high-order bits.  Note that if all states $|\psi_x\>$ have labels $x$ with a common factor---say, $2^j|x$---then we can view the labels as elements of $\Z_{2^{n-j}}$ and apply
\alg{dcomb}
to affect the $n-j$ most significant bits.
Specifically, to make the state $|\psi_{2^j}\>$, we apply \alg{sieve} using \alg{lsbcomb} as the combination procedure that produces states from $S_i$ using states from $S_{i-1}$ for $i \in \{1,\ldots,m_1+1\}$, and \alg{dcomb} (on the $n-j$ most significant bits) as the combination procedure for $i \in \{m_1+2,\ldots,m_1+m_2+1\}$, taking
\[
  S_i=\begin{cases}
  2^{(k-1)i}\{0,1,\ldots,2^{n-(k-1)i}-1\} & \text{for $i \in \{0,1,\ldots,m_1\}$} \\
  2^j\{0,1,\ldots,B_i-1\} & \text{for $i \in \{m_1+1,\ldots,m_1+m_2+1\}$} \\
\end{cases}
\]
where now
\begin{align*}
  B_i &= \floor{2^{n-j}/\rho^{i-m_1-1}} &
  m_1 &= \floor{j/(k-1)} \\
  \rho &= 2^{(n-j-1)/m_2} &
  m_2 &= \Ceil{\frac{n-j}{k-\log_2 2k}}
\end{align*}
and again $k=\floor{\sqrt{\frac{1}{2} \log_2 N \log_2\log_2 N}}$.
When making states in $S_i$ from states in $S_{i-1}$ for $i \in \{1,\ldots,m_1\}$, we cancel $k-1$ bits with $k$ states, so the condition of Lemma \ref{lem:lsbcomb} is satisfied.  For $i=m_1+1$, we cancel $j-(k-1)m_1 = j - (k-1)\floor{j/(k-1)} \le j - (k-1)[j/(k-1) - 1] = k-1$ bits, so again the condition of Lemma \ref{lem:lsbcomb} is satisfied.
For $i \in \{m_1+2,\ldots,m_1+m_2+1\}$, \lem{sievesizesd} implies that the conditions of \lem{dcomb} are satisfied provided $2^{n-j} \ge N_0$.  (If $2^{n-j} < N_0$ then we only need to perform the first $m_1+1$ stages of the sieve, producing a state uniformly at random from $S_{m_1+1}$; in this case $|S_{m_1+1}| = O(1)$, so $O(1)$ repetitions suffice to produce a copy of $|\psi_{2^j}\>$.)
Finally, since $(m_1+m_2+1)k = (1+o(1))n$, the discussion following \lem{sieve} shows that \alg{sieve} takes time $L_N(\frac12,\sqrt2)$.

So far we have covered the case where the group is $A=\Z_N$ with $N$ either odd or a power of $2$.  Now consider the case of a general finite abelian group $A = \Z_{N_1} \times \cdots \times \Z_{N_t}$.  By the Chinese remainder theorem, we can assume without loss of generality that each $N_i$ is either odd or a power of $2$.
Consider what happens if we apply \alg{dcomb} or \alg{lsbcomb} to one component of a product of cyclic groups.  Suppose we combine $k$ states of the form of Eqn.~(\ref{eq:hsqubit}).  For each $i \in \{1,\ldots,k\}$, let $x_i \in \Z_{N_1} \times \cdots \times \Z_{N_t}$ denote the label of the $i$th state, with $x_{i,j} \in \Z_{N_j}$ for $j \in \{1,\ldots,t\}$.  To address the $\ell$th component of $A$, the combination procedure prepares a state
\begin{align*}
  \frac{1}{\sqrt{2^k}} \sum_{y \in \{0,1\}^k} \exp\left(2\pi i \sum_{i=1}^k \sum_{j=1}^t \frac{y_i x_{i,j} s_j}{N_j}\right)|y\>|h(\textstyle \sum_{i=1}^k x_{i,\ell} y_i)\>
\end{align*}
for some function $h$ (a quotient in \alg{dcomb} or a remainder in \alg{lsbcomb}).  For $j \ne \ell$, if $x_{i,j}=0$ for all $i \in \{1,\ldots,k\}$ then $x_j' = \sum_{i=1}^k x_{i,j} (y^\varstar_i-y^\star_i) = 0$, so components that are initially zero remain zero.
Thus, if we can prepare states $|\psi_x\>$ with $x_{\ell} \inrand \Z_{N_\ell}$ (for any desired $\ell \in \{1,\ldots,t\}$) and all other components zero, we effectively reduce the problem to the cyclic case.

To prepare such states, we use a new combination procedure, \alg{zcomb}.  Without loss of generality, our goal is to zero out the first $t-1$ components, leaving the last one uniformly random from $\Z_{N_t}$.  \alg{zcomb} is similar to \alg{dcomb}, viewing the first $t-1$ components of the label $x_i \in \Z_{N_1} \times \cdots \times \Z_{N_t}$ as the mixed-radix integer
\[
  \mixrad{x_i} \defeq \sum_{j=1}^{t-1} x_{i,j} \prod_{j'=1}^{j-1} N_{j'}.
\]
Because we are merely trying to zero out certain components, we no longer require uniformity of the states output by the sieve, which simplifies the procedure and its analysis.

\begin{algorithm}[t]
\caption{Combining non-cyclic states to reduce undesired components}
\label{alg:zcomb}
\begin{algorithmic}[1]
  \REQUIRE Parameters $B,B'$ and states $|\psi_{x_1}\>,\ldots,|\psi_{x_k}\>$ with known $x_1,\ldots,x_k \in \Z_{N_1} \times \cdots \times \Z_{N_t}$ satisfying $\mu(x_i) \in \{0,1,\ldots,B-1\}$ for each $i \in \{1,\ldots,k\}$, with $x_{i,t} \inrand \Z_{N_t}$
  \ENSURE State $|\psi_{x'}\>$ with known $x' \in \Z_{N_1} \times \cdots \times \Z_{N_t}$ satisfying $\mu(x') \in \{0,1,\ldots,B'-1\}$, with $x'_t \inrand \Z_{N_t}$
  \STATE Introduce an ancilla register and compute
  \[
  \frac{1}{\sqrt{2^k}} \sum_{y \in \{0,1\}^k} \exp\left(2\pi i \sum_{i=1}^k \sum_{j=1}^t \frac{y_i x_{i,j} s_j}{N_j}\right)|y\>|\floor{\textstyle\sum_{i=1}^k \mu(x_i) y_i/B'}\>
  \]
  \STATE Measure the ancilla register, giving an outcome $q$ and a state
  \[
    \frac{1}{\sqrt \nu} \sum_{j=1}^\nu \exp\left(2\pi i \sum_{i=1}^k \sum_{j=1}^t \frac{y_i x_{i,j} s_j}{N_j}\right) |y^j\>
  \]
  where $y^1,\ldots,y^\nu \ne 0^k$ are the $k$-bit strings such that $\floor{\sum_{i=1}^k \mu(x_i) y_i^j/B'}=q$
  \STATE\label{step:zcomb-subsetsum} Compute $y^1,\ldots,y^\nu$ by brute force
  \IF{$\nu=1$}
    \STATE Abort
  \ENDIF
  \STATE\label{step:zcomb-project} Project onto $\spn\{|y^1\>,|y^2\>\}$ or $\spn\{|y^3\>,|y^4\>\}$ or \ldots\ or $\spn\{|y^{2\floor{\nu/2}-1}\>,|y^{2\floor{\nu/2}}\>\}$, giving an outcome $\spn\{|y^\star\>,|y^\varstar\>\}$
  \STATE\label{step:zcomb-final} Relabel $|y^\star\> \mapsto |0\>$ and $|y^\varstar\> \mapsto |1\>$ where $\sum_{i=1}^k \mu(x_i) y^\varstar_i \ge \sum_{i=1}^k \mu(x_i) y^\star_i$ WLOG, giving a state $|\psi_{x'}\>$ with $x'_j = \sum_{i=1}^k x_{i,j} (y^\varstar_i-y^\star_i)$ for each $j \in \{1,\ldots,t\}$
\end{algorithmic}
\end{algorithm}

\begin{lemma}\label{lem:zcomb}
\alg{zcomb} runs in time $2^k \poly(\log N)$ and succeeds with probability $\Omega(1)$ provided $B/B' \le 2^k/2k$.
\end{lemma}

\begin{proof}
As in \lem{dcomb} and \lem{lsbcomb},
the running time is dominated by the brute force calculation in \step{zcomb-subsetsum} and the projection in \step{zcomb-project}, both of which can be performed in time $2^k \poly(\log N)$.

We claim that the algorithm is correct if it reaches \step{zcomb-final}.
Since $\sum_{i=1}^k \mu(x_i) y^j_i = q B' + r^j$ where $q$ is independent of $j$ and $0 \le r^j < B'$, we have 
\begin{align*}
  \mu(x')
  = \sum_{j=1}^{t-1} x'_j \prod_{j'=1}^{j-1} N_{j'}
  = \sum_{j=1}^{t-1} \sum_{i=1}^k x_{i,j} (y^\varstar_i - y^\star_i) \prod_{j'=1}^{j-1} N_{j'}
  = \sum_{i=1}^k \mu(x_i) (y^\varstar_i - y^\star_i)
  = r^\varstar-r^\star
  < B'
\end{align*}
as required.  Since $y^\star \ne y^\varstar$ and the $x_{i,t}$ are uniformly random, $x'_t = \sum_{i=1}^k x_{i,t} (y^\varstar_i - y^\star_i)$ is uniformly random as required.

The projection in \step{zcomb-project} fails with probability at most $1/3 + o(1)$.  We claim the algorithm reaches \step{zcomb-project} with probability $\Omega(1)$.
To show this, we need to upper bound the probability that $\nu=1$.  Call a value of $q$ bad if $\nu=1$.  Since $0 \le \sum_{i=1}^k \mu(x_i) y_i \le k(B-1)$, there are at most $kB/B'$ possible values of $q$, and in particular, there can be at most $kB/B'$ bad values of $q$.  Since the probability of any particular bad $q$ is $1/2^k$, the probability that $q$ is bad is at most $kB/B'2^k \le 1/2$.  This completes the proof.
\end{proof}

To apply \alg{zcomb} as the combination procedure for \alg{sieve}, we require a straightforward variant of \lem{sievesizesd}, as follows.

\begin{lemma}\label{lem:sievesizesz}
There is a constant $N_0'$ such that for all $N>N_0'$, letting $B_i = \floor{N /\rho^i}$ where $\rho = N^{1/m}$ and
\begin{align*}
k &= \Floor{\sqrt{\tfrac{1}{2} \log_2 N \log_2\log_2 N}} &
m &= \Ceil{\frac{\log_2 N}{k-\log_2 4k}}
   = \Theta\left(\sqrt{\frac{\log_2 N}{\log_2\log_2 N}}\right),
\end{align*}
we have $B_0=N$, $B_m=1$, and $B_{i-1}/B_i \le 2^k/2k$ for all $i\in \{1,\ldots,m\}$.
\end{lemma}

\begin{proof}
Clearly $B_0=N$, and the value of $\rho$ is chosen so that $B_m = 1$.

We have
$
  B_{m-1} \le N/\rho^{m-1} = \rho
$,
and since
\begin{align*}
  \rho
  &\le N^{\frac{k-\log_2 4k}{\log_2 N}}
  = \frac{2^k}{4k},
\end{align*}
the claimed inequality holds for $i=m$.

For $i \in \{1,\ldots,m-1\}$,
\begin{align*}
  \frac{B_{i-1}}{B_i}
  = \frac{\floor{N/\rho^{i-1}}}{\floor{N/\rho^i}}
  \le \frac{N/\rho^{i-1}}{N/\rho^i - 1}
  = \frac{\rho}{1-\rho^i/N}.
\end{align*}
Since $\rho^i/N \le \rho^{m-1}/N = 1/\rho$, we have $B_{i-1}/B_i \le \rho/(1-1/\rho)$.  Then using
\begin{align*}
  \rho
  = N^{\Theta(\sqrt{\log_2\log_2 N/\log_2 N})}
  = 2^{\Theta(\sqrt{\log_2 N \log_2\log_2 N})},
\end{align*}
we have $\rho \ge 2$ provided $N > N_0'$ for some constant $N_0'$, which implies $B_{i-1}/B_i \le 2^k/2k$.  This completes the proof.
\end{proof}

Combining these ideas, the overall procedure is presented in \alg{hiddenshift}.

\begin{algorithm}[t]
\caption{Abelian hidden shift problem}
\label{alg:hiddenshift}
\begin{algorithmic}[1]
  \REQUIRE Black box for the hidden shift problem in an abelian group $A$
  \ENSURE Hidden shift $s$
  \STATE\label{step:hs-decompose} Write $A = \Z_{N_1} \times \cdots \times \Z_{N_t}$ where each $N_i$ is either odd or a power of $2$
  \FORALL{$i \in \{1,\ldots,t\}$}\label{step:hs-outerfor}
    \IF{$N_i$ is odd}
      \FORALL{$j \in \{0,\ldots,\floor{\log_2 N_i}\}$}\label{step:hs-innerfor-odd}
        \STATE\label{step:hs-sieve-odd}Apply \alg{sieve}, first using \alg{zcomb} to zero out all components except the $i$th one and then using \alg{dcomb} under the $\Z_{N_i}$-automorphism $x \mapsto 2^{-j} x$ to produce a copy of $|\psi_{(0,\ldots,0,2^j,0,\ldots,0)}\>$
        (see the proof of \thm{hiddenshift-alg} for detailed parameters)
      \ENDFOR
    \ELSE
      \STATE Let $N_i = 2^n$
      \FORALL{$j \in \{0,\ldots,n-1\}$}\label{step:hs-innerfor-even}
        \STATE\label{step:hs-sieve-even}Apply \alg{sieve}, first using \alg{zcomb} to zero out all components except the $i$th one, then using \alg{lsbcomb} to make states $|\psi_{(0,\ldots,0,x,0,\ldots,0)}\>$ with $2^j|x$, and finally using \alg{dcomb} to produce a copy of $|\psi_{(0,\ldots,0,2^j,0,\ldots,0)}\>$
        (see the proof of \thm{hiddenshift-alg} for detailed parameters)
      \ENDFOR
    \ENDIF
    \STATE\label{step:hs-reconstruct} Apply \lem{qftreconstruct} with $N=N_i$ to give $s_i$
  \ENDFOR
  \STATE\label{step:hs-output} Output $s=(s_1,\ldots,s_t)$
\end{algorithmic}
\end{algorithm}

\begin{theorem}\label{thm:hiddenshift-alg}
\alg{hiddenshift} runs in time $L_{|A|}(\frac12,\sqrt2)$.
\end{theorem}

\begin{proof}
In \step{hs-decompose}, if the structure of the group is not initially known, it can be determined in polynomial time using \cite{mosca}.  Given the structure of the group, for each term $\Z_{N}$ we can easily factor $N=2^n M$ where $M$ is odd; then $\Z_{N} \cong \Z_{2^n} \times \Z_M$, and we obtain a decomposition of the desired form.

Now suppose without loss of generality that we are trying to determine $s_t$ (i.e., $i=t$ in \step{hs-outerfor}).
The main contribution to the running time comes from the sieves in \step{hs-sieve-odd} (for $N_t$ odd) and \step{hs-sieve-even} (for $N_t$ a power of $2$).

First suppose that $N_t$ is odd.
It suffices to handle the case where $j=0$, so we are making the state $|\psi_{(0,\ldots,0,1)}\>$.
Then we apply \alg{sieve} with
\[
  S_i=\begin{cases}
  \{x \in A: \mu(x) < B_i\} & \text{for $i \in \{0,1,\ldots,m_1\}$} \\
  \{x \in A: \mu(x)=0 \text{~and~} x_t < B_i\} & \text{for $i \in \{m_1+1,\ldots,m_2\}$}
\end{cases}
\]
where
\[
  B_i=\begin{cases}
  \floor{(N/N_t)/\rho_1^i} & \text{for $i \in \{0,1,\ldots,m_1\}$} \\
  \floor{N_t/\rho_2^{i-m_1}} & \text{for $i \in \{m_1+1,\ldots,m_1+m_2\}$}
\end{cases}
\]
with
\begin{align*}
  \rho_1 &= (N/N_t)^{1/m_1} &
   m_1 &= \Ceil{\frac{\log_2 N/N_t}{k-\log_2 4k}} \\
  \rho_2 &= (N_t/2)^{1/m_2} &
  m_2 &= \Ceil{\frac{\log_2 N_t/2}{k-\log_2 2k}}
\end{align*}
and $k = \floor{\sqrt{\frac{1}{2} \log_2 N \log_2\log_2 N}}$.
We use \alg{zcomb} as the combination procedure for the first $m_1$ stages of \alg{sieve}.  By \lem{sievesizesz}, the condition of \lem{zcomb} is satisfied provided $N/N_t > N_0'$; otherwise we can produce a state with a label from $S_{m_1}$ in only $O(1)$ trials.  Then we proceed to apply \alg{dcomb} as the combination procedure for the remaining $m_2$ stages of \alg{sieve}.  By \lem{sievesizesd}, the conditions of \lem{dcomb} are satisfied provided $N_t > N_0$; otherwise, producing states with labels from $S_{m_1}$ already suffices to produce the desired state with constant probability.
Since $(m_1+m_2)k = (1+o(1))\log_2 N$, \step{hs-sieve-odd} takes time $L_{|A|}(\frac12,\sqrt2)$ (see the discussion following the proof of \lem{sieve}).

Now suppose that $N_t = 2^n$ is a power of $2$.  Then we apply \alg{sieve} with
\[
  S_i=\begin{cases}
  \{x \in A: \mu(x) < B_i\} & \text{for $i \in \{0,1,\ldots,m_1\}$} \\
  \big\{x \in A: \mu(x)=0 \text{~and~} x_t \in 2^{(k-1)i}\{0,1,\ldots,2^{n-(k-1)i}\}\big\} & \text{for $i \in \{m_1+1,\ldots,m_1+m_2\}$} \\
  \big\{x \in A: \mu(x)=0 \text{~and~} x_t \in 2^j\{0,1,\ldots,B_i-1\}\big\}
  &\begin{array}{@{}r@{}l@{}}\text{for~} i \in \{& m_1+m_2+1,\ldots, \\[-3pt] & m_1+m_2+m_3+1\}\end{array}
\end{cases}
\]
where
\[
  B_i=\begin{cases}
  \floor{(N/N_t)/\rho_1^i} & \text{for $i \in \{0,1,\ldots,m_1\}$} \\
  \floor{2^{n-j}/\rho_3^{i-m_1-m_2-1}} & \text{for $i \in \{m_1+m_2+1,\ldots,m_1+m_2+m_3+1\}$}
\end{cases}
\]
with
\begin{align*}
  \rho_1 &= (N/N_t)^{1/m_1} &
   m_1 &= \Ceil{\frac{\log_2 N/N_t}{k-\log_2 4k}} \\
  & & m_2 &= \floor{j/(k-1)} \\
  \rho_3 &= 2^{(n-j-1)/m_3} &
  m_3 &= \Ceil{\frac{n-j-1}{k-\log_2 2k}}
\end{align*}
and again $k = \floor{\sqrt{\frac{1}{2} \log_2 N \log_2\log_2 N}}$.
We use \alg{zcomb} as the combination procedure for the first $m_1$ stages, \alg{lsbcomb} for the next $m_2+1$ stages, and \alg{dcomb} (on the $n-j$ most significant bits) for the final $m_3$ stages.
By \lem{sievesizesz} and \lem{sievesizesd}, the conditions of \lem{zcomb} and \lem{dcomb} are satisfied, respectively.  Since we cancel at most $k-1$ bits in each stage that uses \alg{lsbcomb}, the conditions of \lem{lsbcomb} are satisfied for the intermediate stages.
Finally, since $(m_1+m_2+m_3+1)k = (1+o(1)) \log_2 N$, \step{hs-sieve-even} takes time $L_{|A|}(\frac12,\sqrt2)$.

The loops in \step{hs-outerfor}, \step{hs-innerfor-odd}, and \step{hs-innerfor-even} only introduce polynomial overhead.
\step{hs-reconstruct} takes polynomial time and
\step{hs-output} is negligible.
Thus the overall running time is $L_{|A|}(\frac12,\sqrt2)$ as claimed.
\end{proof}

\end{document}